\documentclass[aps,10pt,groupedaddress,twocolumn,superscriptaddress,nofootinbib,prl]{revtex4-1}

\usepackage{graphicx}
\usepackage{dcolumn}
\usepackage{bm}
\usepackage{hyperref}
\usepackage{natbib}

\usepackage{float}
\usepackage{graphicx,subfigure}
\usepackage{booktabs}
\usepackage{glossaries}
\usepackage{multirow}

\newacronym{wtn}{ITN}{International Tourism Network}
\newacronym{wan}{WAN}{World Airline Network}
\newacronym{unwto}{UNWTO}{World Tourism Organization}
\newacronym{si}{SI}{Supplementary Information}

\begin{document}

\title{Impact of Inter-Country Distances on International Tourism}

\author{T. Verma \thanks{Correspondence and requests for materials should be addressed to T. V. (vtrivik@ethz.ch)}} 
\affiliation{Institute for Terrestrial Ecosystems, ETH Z\"urich, Universit{\"a}tstrasse 16, 8092 Z\"urich,
  Switzerland}
 \affiliation{Faculty of Technology, Policy and Management, Delft University of Technology, 2628BX Delft, the Netherlands}
  
\author{L. Rebelo} 
\affiliation{Departamento de F\'isica, Faculdade de Ci\^encias, Universidade de Lisboa, P-1749-016 Lisboa, Portugal}
\affiliation{Centro de F\'isica Te\'orica e Computacional, Universidade de Lisboa, 1749-016 Lisboa, Portugal}
  
\author{N. A. M. Ara\'ujo} 
\affiliation{Departamento de F\'isica, Faculdade de Ci\^encias, Universidade de Lisboa, P-1749-016 Lisboa, Portugal}
\affiliation{Centro de F\'isica Te\'orica e Computacional, Universidade de Lisboa, 1749-016 Lisboa, Portugal}

\begin{abstract}
Tourism is a worldwide practice with international tourism revenues increasing from US$\$ 495$ billion in $2000$ to US$\$ 1340$ billion in $2017$. Its relevance to the economy of many countries is obvious. Even though the \gls{wan} is global and has a peculiar construction, the \gls{wtn} is very similar to a random network and barely global in its reach. To understand the impact of global distances on local flows, we map the flow of tourists around the world onto a complex network and study its topological and dynamical balance. We find that although, the \gls{wan} serves as infrastructural support for the \gls{wtn}, the flow of tourism does not correlate strongly with the extent of flight connections worldwide. Instead, unidirectional flows appear locally forming communities that shed light on global travelling behavior inasmuch as there is only a $15\%$ probability of finding bidirectional tourism between a pair of countries. We conjecture that this is a consequence of one-way cyclic tourism by analyzing the triangles that are formed by the network of flows in the \gls{wtn}. Finally, we find that most tourists travel to neighbouring countries and mainly cover larger distances when there is a direct flight, irrespective of the time it takes. 
\end{abstract}

\pacs{Valid PACS appear here}
\maketitle

\section*{\label{sec:Intro}Introduction}
Mobility has always been a presence in the life of human beings, starting with the first \textit{Homo erectus} who began to disperse from Africa soon after their emergence in a journey for survival \cite{Carotenuto2016,harari2015sapiens}. Historically, it paved the way for human exploration and growth: settlements came up, villages became towns, and towns turned into cities \cite{Idyorough1998}. Nowadays, human travel is primarily about business and tourism and relies on transportation networks which are either local, regional or global. 

The World Airline Network is arguably the largest global transportation network, serving business, tourism and cargo needs of the world. In the last decade, \gls{wan} has impacted tourism with revenues increasing from US$\$ 495$ billion in $2000$ to US$\$ 1340$ billion in $2017$, while international tourist arrivals grew from $674$ to $1322$ million in this time span and are expected to reach $1.8$ billion by the year $2030$, according to industry projections \cite{unwto_th}. However, the magnanimity of the economic and infrastructural scales of this network does not put forth any understanding of how tourists perceive distances: how far do people really travel? Is tourism a global phenomenon? What is our predilection for the destinations we visit?

The representation of the \gls{wtn} as a complex network allows us to use methods and insights from network theory and data science to investigate the flow of international tourism in a quantitative manner. Simultaneously, this formulation also aids in correlating the structural differences of the tourism network with an omnipresent infrastructure network like the \gls{wan} that can, in principle, enable one to travel anywhere in the world. A large number of real world systems, from social to biological and infrastructural to financial \cite{metabolic_pathways,food_webs_review,taxonomy,email_network,financial_network}, have also been studied successfully using network theory to classify their structure and dynamic behavior \cite{sd_brain,sd_biology,boccaletti_structure_dynamics}. 

In this article, we start with a coarse-grained version of the \gls{wan} at the country level, i.e. each airport of a country is aggregated to one node and links are aggregated between countries (all calculations with the \gls{wan} refer to its original state and simplification to the country level has only been done to establish the \gls{wtn}). Then, we consider the countries present in the \gls{wan} as nodes of the \gls{wtn} and add a link between them if there is flow of tourists from one country to another. In the simplified \gls{wan} if a flight exists from country $A$ to $B$, so must exist a connection in the reverse direction and through the same path. In the \gls{wtn}, this is not the case as tourism is not necessarily a reciprocal phenomenon. We classify the links by the reciprocity between their attaching nodes, i.e., if there is a two-way connection between the nodes, the link joining them is undirected but, if the direction of the connection is relevant, they are called directed. Furthermore, we weigh the links using flow of tourists along the direction. We study the balance of the flow of tourism and analyze the impact of geographical and chemical distances (number of connections between two countries using the \gls{wan}) on tourism. 

\begin{figure}
\centering
\includegraphics[width=0.4\textwidth]{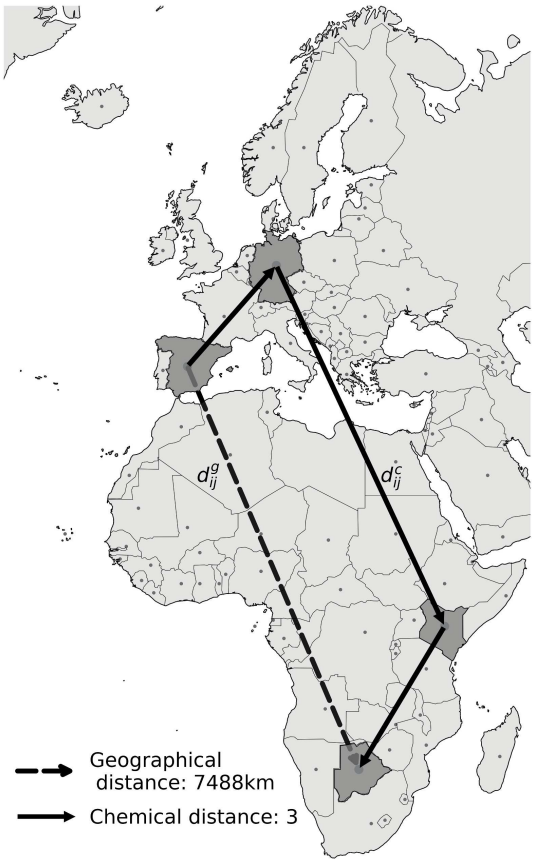}
\caption{\textbf{Representation of the \gls{wtn}}: Illustration of the tourism flow from Spain to Botswana, for which the geographical distance $d^g_{ij}$ is $7488$ km and the chemical distance in the coarse-grained \gls{wan} $d^c_{ij}$ is $3$, illustrating that tourists going from Spain to Botswana have to take at least two inter-country connections, travelling, for example, through Germany and Kenya, which is a much larger geographical distance than what separates the two countries.}
\label{fig:wtnabstraction}
\end{figure}

\section*{\label{results}Results}
\begin{figure*}
\subfigure[Weighted]{
\includegraphics[width=0.45\textwidth]{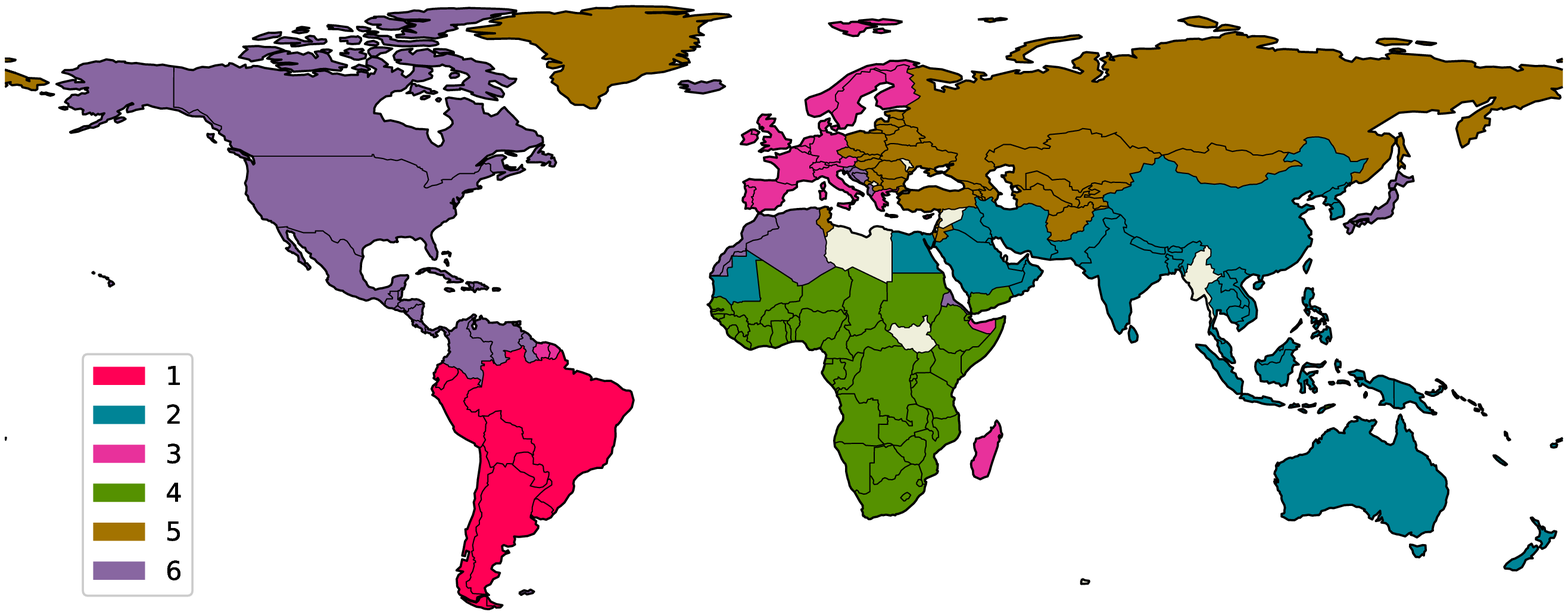}
\label{fig:commWv1}}
\subfigure[Normalized]{
\includegraphics[width=0.45\textwidth]{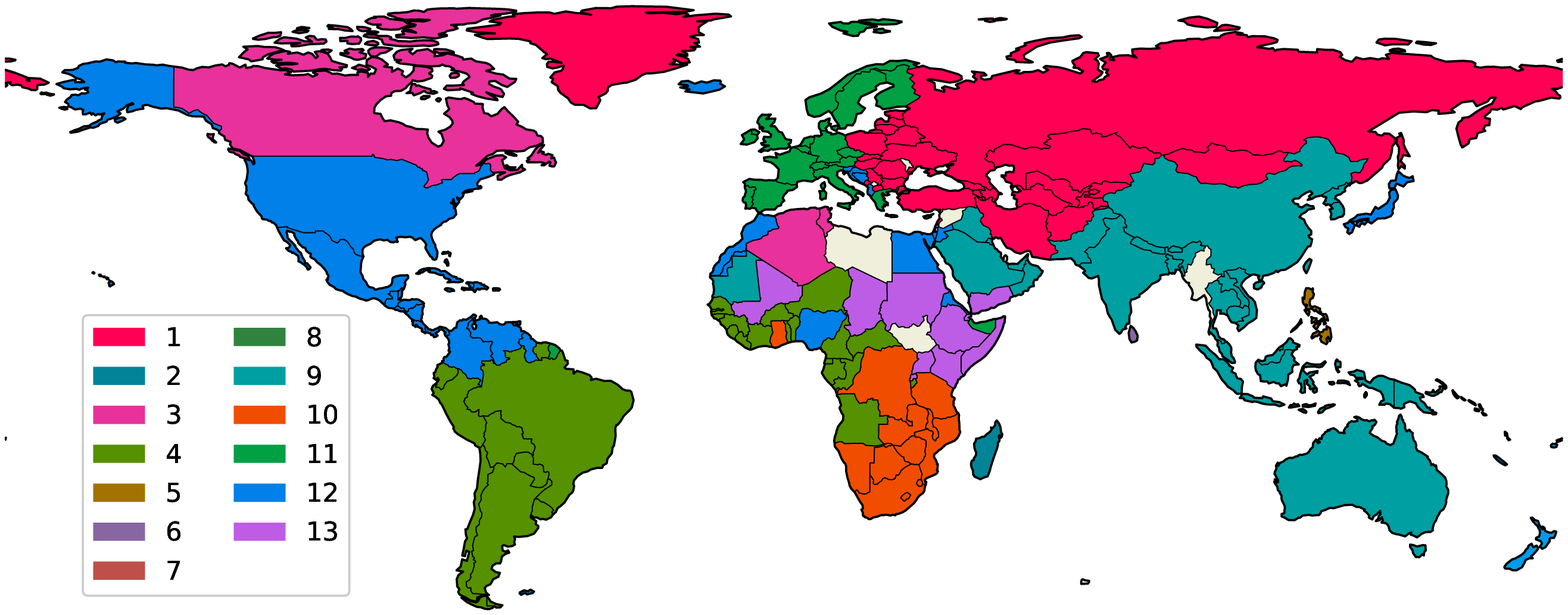}
\label{fig:commWv2}}
\caption{(a) \textbf{Weighted Community Structure of the \gls{wtn}}: Links are weighted as flow of tourists. The number of communities is six and the modularity value is $Q = 0.54$. (b) \textbf{Normalized Community Structure of the \gls{wtn}}: Links are weighted as flow of tourists divided by the population of the sending country. The number of communities is $13$ and the modularity is $Q = 0.7$.}
\end{figure*}

The \gls{wtn} comprises of $N = 214$ nodes (countries) and $L = 4148$ directed links (flow of tourists), with a sparse link density of
\begin{equation}
\frac{L}{N(N-1)} = 0.091, 
\end{equation}
with only $9.1\%$ of directed pairs of countries in the network enabling tourism. The distribution of the \textit{out-degree} $k^{out}_i$ for the \gls{wtn} is remarkably different from the physical infrastructure of the \gls{wan} that enables tourism in the first place. Note in the \gls{si}˜\ref{supp} that the \gls{wtn} is heterogeneous like an Erd{\"o}s-R{\'e}yni graph \cite{Erdos} while the \gls{wan} follows a truncated power-law distribution with a decay function \cite{trivikStructure}. The average number of countries visited by residents of a country is given by the average out-degree,
\begin{equation}
<k^{out}> = \frac{\sum_{i}^{N}{k^{out}_i}}{N} = 19.38,
\end{equation}
with a standard deviation value of $\sigma_{<k^{out}>} = 11.67$. Since all links in the \gls{wtn} have a start and end node, the total in and out-degree are the same, as are their averages. The average out-degree value indicates that a lesser number of countries (when compared to the number of nodes in the network) are readily experiencing tourism and its standard deviation reflects the disparity in varied tourism patterns of countries. While the average degrees are the same, a much larger deviation in the in-degree of $\sigma_{<k^{in}>} = 45.89$ explains how tourists originate from many countries around the world but travel to only a handful. A careful investigation shows that this result stems from the existence of isolated territories that are typically clustered with only a handful of neighboring countries where the cost of flying is low (see Fig. \ref{fig:commWv2}). We can clearly map the six communities to different regions of our planet using tourism flow as weights using the standard community detection method: Western Europe, Eastern Europe, Middle and Far-East with Oceania, South America, North America and Africa (see Fig. \ref{fig:commWv1}) \cite{newman-modularity}.

Even though the communities presented in Fig. \ref{fig:commWv1} are a close approximation to the geopolitical division of our planet, every country has a different scale of outgoing tourism. Considering this, we analyze the communities by using re-scaled weights of tourism flow by population of the originating country. The result is shown in Fig. \ref{fig:commWv2}. Our analysis shows that people prefer to travel to countries within their continent or sub-continent, even though the effects of globalization have created an intricate network of airlines around the world with $68.08\%$ of intercontinental connections.

In the \gls{wtn}, we define the shortest path length from $i$ to $j$ as the minimal distance between the nodes $d^c_{ij}$ (see Fig. \ref{fig:wtnabstraction}) which is the number of flight connections required to travel from one country to another over the coarse-grained \gls{wan}. We observed that the average shortest path length required by a resident of one country to travel to another is $<l> = 2.32$ with a standard deviation of $\sigma_{<l>} = 0.69$. The distribution of the path lengths for the pairs of nodes in the \gls{wtn} is shown in the \gls{si}˜\ref{supp}. This result means that, on average, a traveller is taking $1.32$ connections in order to go from their country to any other. It is a smaller value than what has been previously reported for the \gls{wan} \cite{trivikStructure}, and in comparison depicts how tourism is not strongly correlated with the availability of the physical infrastructure in the \gls{wan}, which may also be used for business and other needs.

\begin{figure}
\centering
\subfigure[Cycle-triangles]{\label{subfig:cycles}\includegraphics[width=0.225\textwidth]{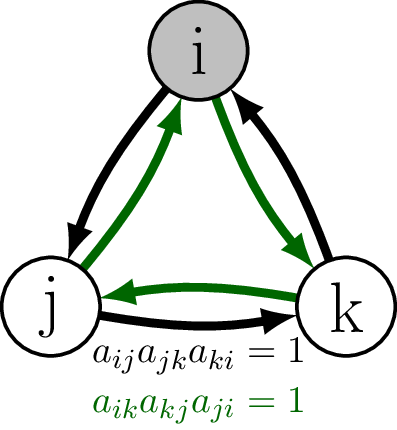}}
\subfigure[In-triangles]{\label{subfig:in-triangles}\includegraphics[width=0.225\textwidth]{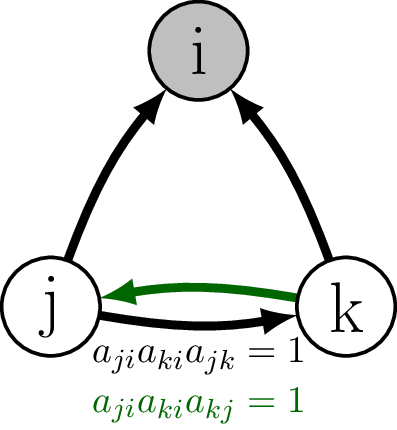}}
\subfigure[Out-triangles]{\label{subfig:out-triangles}\includegraphics[width=0.225\textwidth]{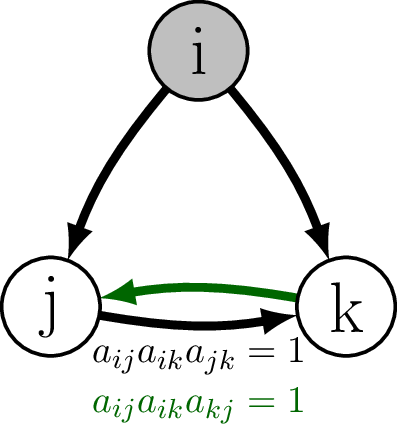}}
\subfigure[Bridge-triangles]{\label{subfig:bridge-triangles}\includegraphics[width=0.225\textwidth]{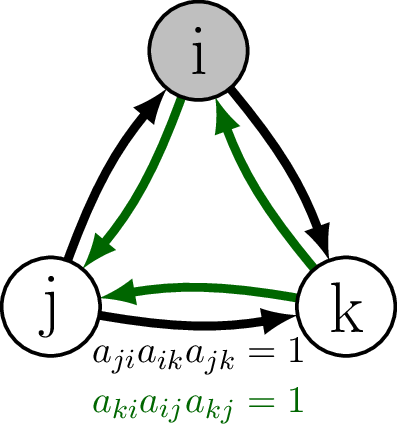}}
\caption{\textbf{Types of Triangles}: Triangles in directed networks and their designations. Each type of triangle corresponds to a different product of the adjacency matrix, which is shown below each sub-figure.}
\label{fig:triangles}
\end{figure}

The \gls{wan} is topologically bidirectional by construction, i.e. there is almost always a return flight available between a pair of airports \cite{Newman:2010:NI:1809753}. By contrast, since the tourism network has an uneven directed flow of tourists from one country to another, only $15\%$ of the links are bidirectional. To quantify the average ratio of tourism flow ($f_{ij}^t$) between nodes, we define the weighted reciprocity as,
\begin{equation}
r^{f} = \frac{1}{L_{\substack{2-way \\ pairs}}} \sum_i \sum_{\substack{j > i \\ j \in n_i^{out} \\ i \in n_j^{out}}} \begin{cases}
\frac{f_{ij}^t}{f_{ji}^t} \qquad \ f_{ij}^t < f_{ji}^t\\
\frac{f_{ji}^t}{f_{ij}^t} \qquad \ f_{ji}^t < f_{ij}^t,
\end{cases}
\end{equation}
and we obtain $r^{f} = 0.44$, meaning that on average, on bidirectional links, the fraction of tourism flow on the lesser used direction is $44\%$ of the flow the other way. This difference could be attributed to the heterogeneity of the economies of the countries represented in the \gls{wtn} and highlights the enormity of an uneven transfer of wealth through tourism. If a traveler moves from their country of residence, they contribute to the economy of the place they travel to, which in turn accumulates and contributes to another country's economy in part if there is movement on that segment of the cycle. We define the flow of tourism wealth using the amount of cyclic clustering that exists in the network for each country. For undirected networks there is only one possible triangle formed between each triplet of nodes. \gls{wtn} is, however, directed in its movement of travelers \cite{fagiolo_clustering}, in which case that number increases to eight, as shown in Fig. \ref{fig:triangles}. We only consider the cycle-triangles (see Fig. \ref{subfig:cycles}) for the calculation of the clustering coefficient since these triangles indicate there is a unidirectional cyclic flow of tourism wealth among countries. Cycle-triangles form cliques of size three, which are triplets of nodes that are all connected among themselves. In a tourism network, these cliques transfer wealth back to themselves, thereby all countries receiving back a portion of their tourism economic value spent abroad.

Taking into account these factors, we introduce the \textit{cyclic clustering coefficient} which is defined as,
\begin{equation}
C_i^{cyc} = \frac{N_i^{cyc}}{n_i^D (n_i^D - 1)} \ ,
\label{eq:cycCC}
\end{equation}
where $N_i^{cyc}$ represents the number of cycle-triangles involving node $i$ and $n_i^D$ represents the number of potential cyclic connections. We expect the probability of occurrence of cycles in a directed network to be of the order of the link density, since if a node is connected to two other nodes, the latter only need to form a link between themselves (in the right direction), in order to form a cycle. The average value of cyclic clustering coefficient for the nodes of the \gls{wtn}  is $<C^{cyc}> \ = 0.015$, which is of the order of the link density, as expected. It is, though, of a small order for a network built on top of the \gls{wan}, which itself has a high value ($C=0.62$ \cite{guimera}).

\begin{figure*}
\subfigure[Distribution]{
\includegraphics[width=0.45\textwidth]{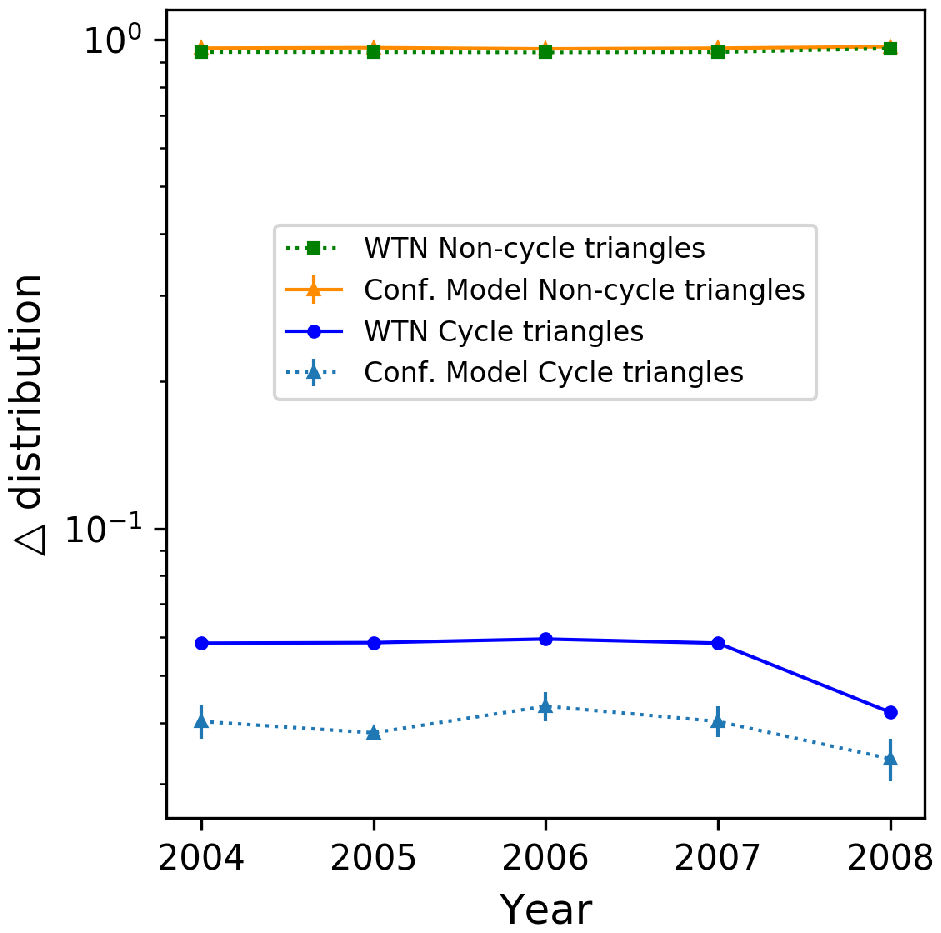}
\label{fig:triDistrComp}}
\subfigure[Participation]{
\includegraphics[width=0.45\textwidth,height=8.075cm]{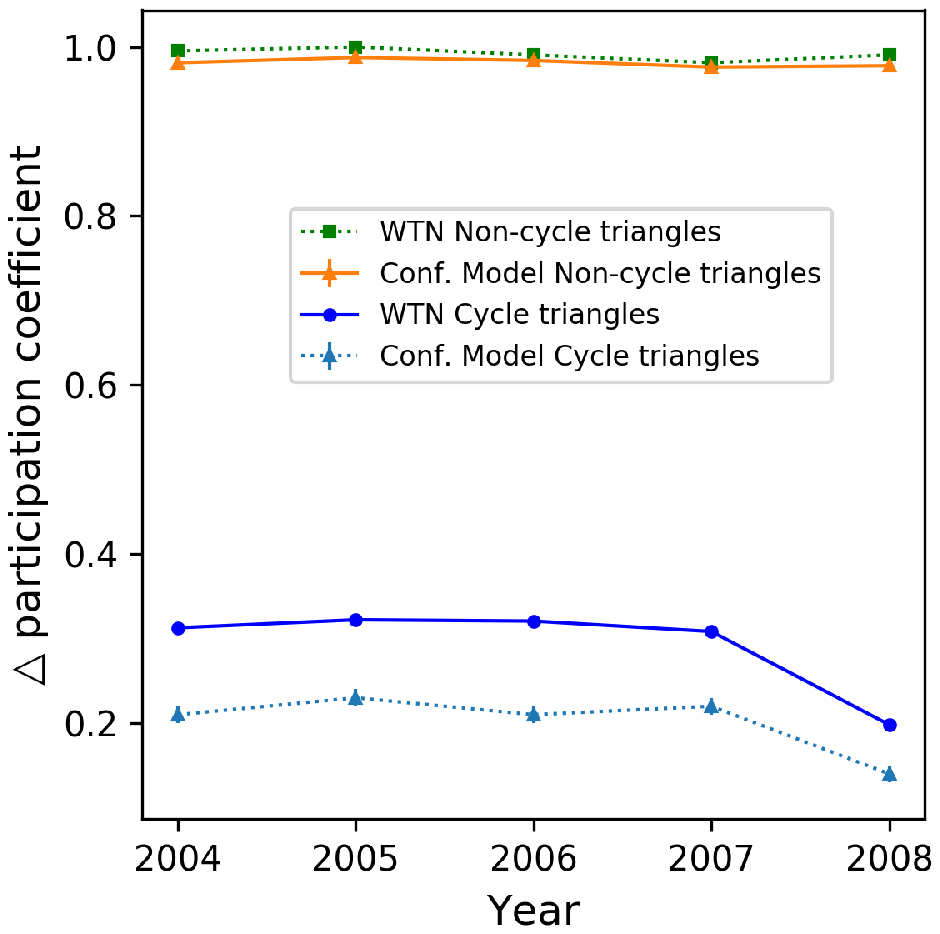}
\label{fig:triPartCoeff}}
\caption{\textbf{Change in distribution and participation of cycles over time}: (a) Triangle distribution values for the \gls{wtn} and the configuration model networks. (b) Triangle participation coefficient values for the \gls{wtn} and the configuration model networks. These networks were generated by randomizing the links present in the \gls{wtn} while keeping the degree distribution of its nodes intact. The results present here are obtained after simulating $100$ instances of these networks and taking the average of each parameter.}
\end{figure*}

The weighted clustering coefficient of the \gls{wtn}, $C_w = \frac{\text{total value of closed triplets}}{\text{total value of triplets}}$, for which a triplet is a trio of nodes which is closed if all nodes are connected with each other \cite{opsahl}, is $C_w = 0.0077$. If we neglect the direction of the links in the \gls{wtn} and calculate merely the unweighted and undirected clustering coefficient \cite{watts_strogatz}, we obtain $<C^{und}> = 0.79$, a much larger value than for the directed version of the network and slightly larger than the \gls{wan} itself. This difference sheds light on the profile of tourism around the world, in that the world is very well connected in the context of air travel with built-in redundancies in connectivity but flow of tourism is mostly unidirectional, exposing the unequal privilege of human mobility in terms of tourism, where people tend to go from richer countries to poorer ones, but not otherwise. 

The cyclic triangle distribution for the \gls{wtn} is about $6\%$. In addition, only $25\%$ of the countries exhibit a cyclic triangle relationship with other countries. In order to have a benchmark, we reproduce the \textit{configuration model} by rewiring all the links in a network in order to generate a new network, which is uncorrelated in terms of node-node correlations, while keeping the degree sequence of the nodes intact \cite{Newman1}. The structural representations of the two networks (see Fig. \ref{fig:triDistrComp} and \ref{fig:triPartCoeff}) indicate that the distribution of triangles in the \gls{wtn} resembles the distribution of triangles within a random network as opposed to a real-world network. Detailed tabular report is presented in the \gls{si}˜\ref{supp}.

\begin{figure*}
\subfigure{
\includegraphics[width=0.45\textwidth]{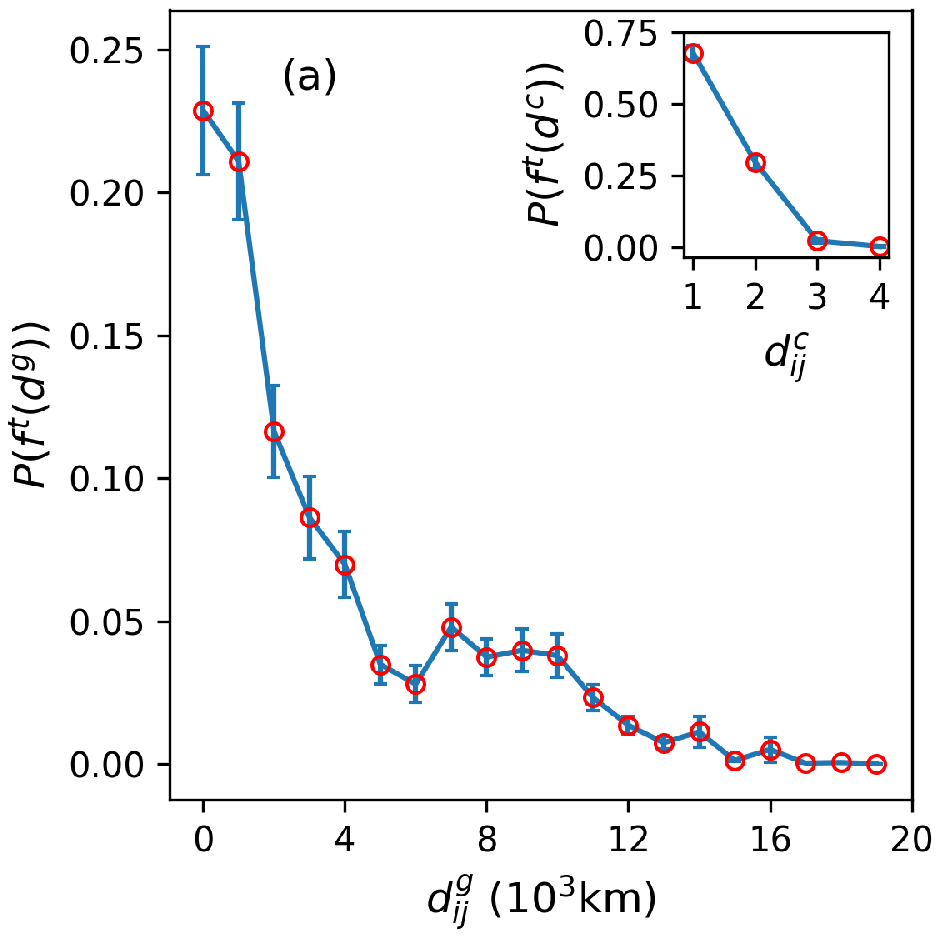}
\label{fig:touristicShare}}
\subfigure{
\includegraphics[width=0.45\textwidth]{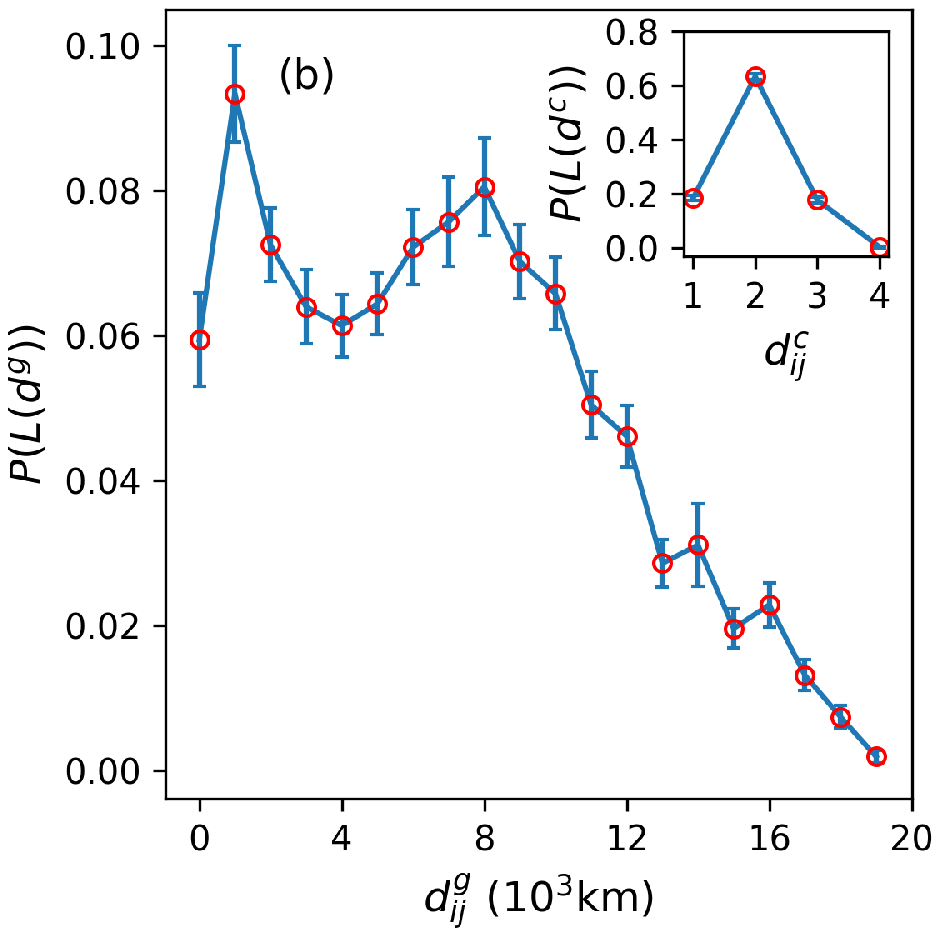}
\label{fig:connectionsShare}}
\caption{\textbf{Distribution of Tourism and Links}: (a) Outgoing tourism flow distribution of the \gls{wtn} using geographical and chemical (inset) distances as spatial indices. (b) Neighbourhood distribution of the \gls{wtn} using geographical and chemical (inset) distances as spatial indices. The countries in the neighbourhood of an originating country are the ones that receive an influx of tourism through direct or indirect links. The error bars represent the standard error.}
\end{figure*}

While the \gls{wtn} is formed on the physical linkages of the \gls{wan}, its spatial organization differs for both chemical and geographical distances (see Methods˜\ref{methods}). In order to study the influence of the \emph{chemical distance} on the number of tourists going from one country to another, we calculate the fraction of tourists as a function of the chemical distance (from $1$, a direct flight, to $4$, the largest chemical distance in the \gls{wtn}). The distribution of country-wide outgoing fraction of tourism flow for all chemical distances ($d_{ij}^c$) is expressed as,
\begin{equation}
f_{ij}^t(d_{ij}^{c} \in [1,4]) = \frac{F_{ij}^t}{\sum \limits_{j \in \{n_i^{out}\}} F_{ij}^t},
\label{eq:outflow}
\end{equation}
where $F_{ij}^t$ is the number of tourists from $i$ to $j$ and $\{n_i^{out}\}$ is the set of countries receiving tourists from $i$.

We measure the distribution of tourists as a function of the geographical distance by dividing the distance between countries into $1000$ km bins. We notice that the peak of fraction of tourists is in the $(0,1000]$ km interval. Note that the first two bins have very similar values, which means that people are almost as likely to travel to a destination that is at most $1000$ km away as they are to a destination between $1000$ and $2000$ km away, and the distribution decays with the distance after. The resulting distribution of the fraction of tourists originating from a country on each link is presented in the inset of Fig. \ref{fig:touristicShare}, where we observe that on average, for the entire network, the fraction of tourists is highest on single link segments to neighbouring countries and decreases monotonically with increase in the number of links between the nodes, thereby indicating that tourists prefer direct connections. Approximately $77\%$ of global tourists travel as far as $4000$ km. Similarly, Fig. \ref{fig:connectionsShare} shows the fraction of direct or indirect neighbours that observe tourists between countries based on geographical distances (again binned by $1000$ km) and chemical distances from originating country.

The distribution of links for chemical distances in the inset of the figure suggests that even though most outgoing tourism links from a country on average are at a distance $d_{ij}^c = 2$, typically a lot more travelers fly to countries that are directly connected for tourism. Geographically speaking, the link distribution does not exhibit a large difference in the values, at least for the $(0,11000]$ km range, as previously observed for the tourism-share, indicating that there is still some variety in choice of travel to farther countries. More importantly, note that while the probability of finding links is not much different as a function of the geographical distance $(0,8000]$ km, the network is utilizing the underlying \gls{wan} only locally exhibited by the peak in the chemical distance distribution. Figures \ref{fig:touristicShare} and \ref{fig:connectionsShare} illustrate the average distributions of tourism and links between countries for various distances. It specifically highlights that most amount of traffic moves on lesser direct links and lesser amount of traffic has a rather large subset of links available due to a variety of travelling behavior. However, the relationship between each individual country's travel preference in terms of the two perceptions of distance (chemical and geographical) is unclear. 
\begin{figure}
\centering
\includegraphics[width=0.45\textwidth]{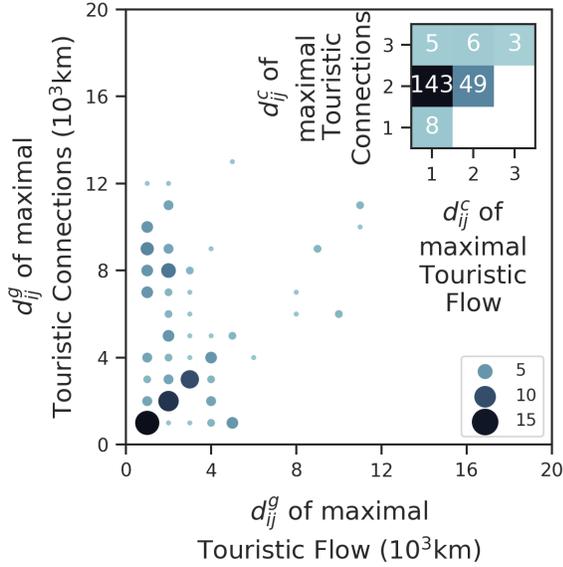}
\caption{\textbf{Variability in supply and demand}: A binned scatter plot indicates the relationship between geographical distances of each country in the \gls{wtn} for which the tourism share and number of links maximizes. The squares without a number correspond to unregistered occurrences. In the inset, a heatmap indicates the relationship between chemical distances for which the tourism share and number of links maximizes.}
\label{fig:flowLinks}
\end{figure}

\begin{figure}
\centering
\includegraphics[width=0.45\textwidth]{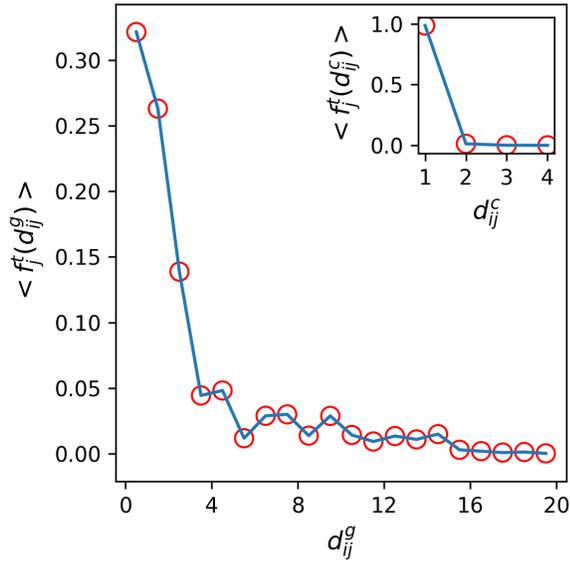}
\caption{\textbf{Distribution of Tourism and Links}: Country-wide incoming distribution of average tourism flow of the \gls{wtn} using geographical and chemical (inset) distances as spatial indices.}
\label{fig:vParam}
\end{figure}

Figure \ref{fig:flowLinks} compares the distances at which the distributions of tourism share and links maximizes for each country. We notice that the demand (tourism flow) and supply (links) variability for geographically binned ($1000$ km) distances do not correlate. Since the average fraction of tourists from any country that prefer a range of $(0,2000]$ km on a direct link is $0.44 \pm 0.02$, there are not enough data points beyond that distance range to report anything statistically significant. However, similar to Eq. \ref{eq:outflow}, the distribution of average incoming flow of tourists ($f_{j}^t$) for all chemical distances ($d_{ij}^c$) is a proxy for the demand being met and is expressed as,
\begin{equation}
f_{j}^t(d_{ij}^{c} \in [1,4]) = \frac{\sum \limits_{i \in \{n_j^{in}\}} F_{ij}^t}{n_j^{in}},
\label{eq:v_param}
\end{equation} 
where {$n_j^{in}$} is the set of in-neighbors of node $j$. Our analysis shows that, on average, a country receives the most number of incoming tourists from a direct connection (see Fig. \ref{fig:vParam}, normalized over all points in the plot). The geographical distribution of incoming tourism also shows that countries on average receive roughly $72\%$ of global tourists from $(0,3000]$ km range.

\begin{figure}
\subfigure[Structural Balance]{
\includegraphics[width=0.47\textwidth]{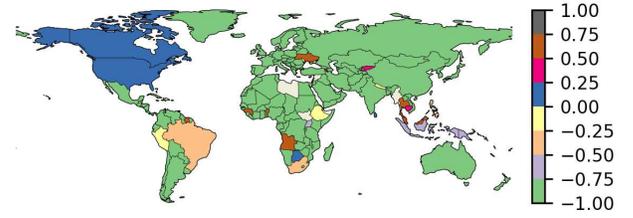}
\label{fig:tParamwtn}}
\subfigure[Functional Balance]{
\includegraphics[width=0.47\textwidth]{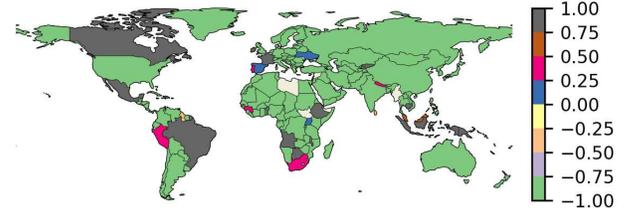}
\label{fig:pParamwtn}
}
\caption{\textbf{Structural and Functional Balance}: Color coded map of the countries in the \gls{wtn} for (a) structural balance $B^s_i$ and (b) functional balance $B^t_i$ . The majority of countries are in the $[-1,-0.75]$ region indicating that most tourism is outgoing in nature. The countries for which we have no data are represented in grey.}
\end{figure}

\begin{figure*}
\subfigure{
\includegraphics[width=0.45\textwidth]{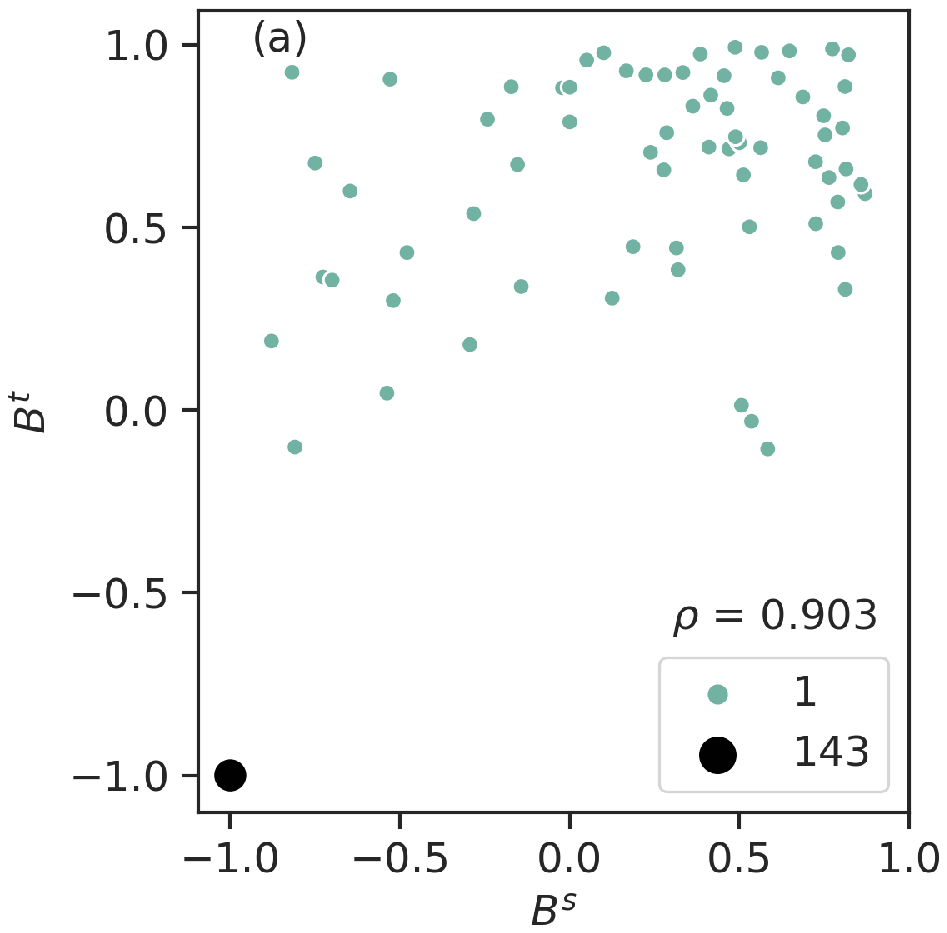}
\label{fig:tpCorrelation}}
\subfigure{
\includegraphics[width=0.45\textwidth]{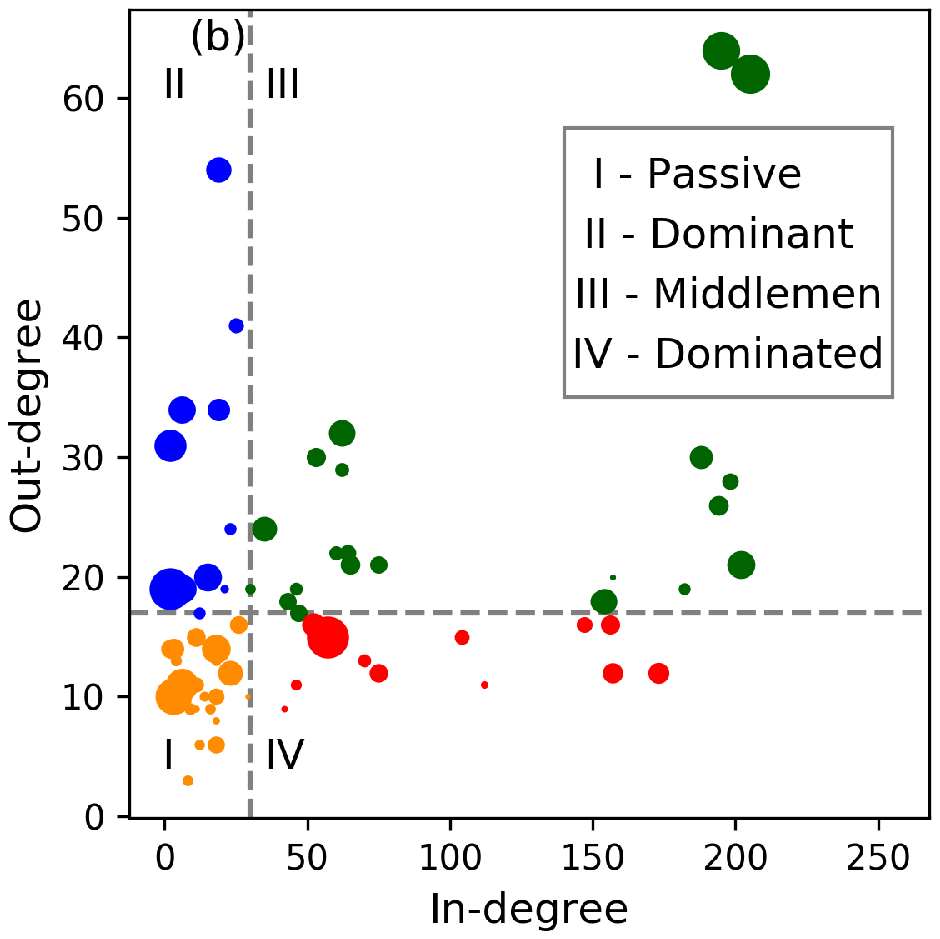}
\label{fig:quantilePlot}}
\caption{\textbf{Categorical Balance}: (a) Scatter plot of the $B ^t$ and $B^s$ parameters. Their correlation coefficient is $\rho=0.903$ which means that countries typically have similar structural and functional balance. (b) $50\%$-Quantile division of the in and out-degree of the countries present in the \gls{wtn} with a color code respective to their strength of bi-directional tourism (I - Passive, II - Dominant, III - Middlemen, IV - Dominated). The radius of each marker is proportional to the ratio of its outgoing and incoming touristic flow. Note: we did not included the nodes with zero in degree in the figure, for the visualization to not have infinite ratio for size.}
\end{figure*}

Certainly, the network is not in balance, i.e. the reciprocity of tourism is not bidirectional and equal. The disparity in the flow distribution however, is not a consequence of the connectivity of regions because these links form the physical infrastructure of the \gls{wan} which is universally reciprocal. So, which nodes in the network are tipping the scales of this balance? A measure of the structural and functional balance of tourists $B^s$ and $B^t$ takes into account the degree and tourist traffic from (to) each node,
\begin{equation}
B^s_i = \frac{k^{in}_i - k^{out}_i}{k^{in}_i + k^{out}_i},
\end{equation}
\begin{equation}
B^t_i = \frac{F^{in}_i - F^{out}_i}{F^{in}_i + F^{out}_i},
\end{equation}
where $F^{in(out)}_i$ is the number of incoming (outgoing) tourists from (to) node $i$. By definition, both $B^s_i$ and $B^t_i \in [-1,1]$. Most countries have a value $-1$, in that there is a strict flow of outgoing tourists only, while $1$ indicates countries with strictly incoming flow. The distribution of the structural and functional balance throughout the network is illustrated in Fig. \ref{fig:tParamwtn} and \ref{fig:pParamwtn}. Note that most countries are in the region of $[-1,-0.75]$ indicating that tourism is in general more outgoing than incoming. Furthermore, the two quantities of balance are correlated with $\rho = 0.903$ as shown in Fig. \ref{fig:tpCorrelation}. Figure \ref{fig:quantilePlot} shows that the four quadrants in the $50\%$-quantile division plot are not equally spaced out; most nodes have a low incoming degree, thereby reinforcing our analysis.

\section*{\label{discussion}Discussion}
In this article, we study the effect of distances on tourism and how perception changes from small to large distances. We observe most people prefer to travel to close by touristic destinations, either in terms of flight connections or in geographical distance. The relationship between the \gls{wan} and the \gls{wtn} sheds light on how the two structures are very dissimilar and one is not molded by the other. In fact, tourism demand through the \gls{wtn} is not always matched by the supply network of the \gls{wan} indicating that business and cargo needs form a large part of the airline network as well. Finally, assessing community structures illustrated that tourism is largely biased since most of the tourism occurs in countries with stronger geopolitical relations and a history with one's country of origin. We believe it is imperative to further explore how wealth is transferred through tourism and design possible optimization strategies for the \gls{wan} in order to meet the demands of the high season of tourism by using a variable underlying airline network.

\section*{\label{methods}Materials and Methods}
The \gls{wtn} data were provided by the \gls{unwto} and consists of the origin and number of ``arrivals of non-resident tourists at national border'' from $2004$ to $2008$ (inclusive) for each country. These overnight visitors are people who travel to a country other than their country of residence, but outside their usual environment, for a period not exceeding $12$ months and whose main purpose in visiting is other than an activity remunerated from within the country visited \cite{unwto_th}. The sources and collection methods differ across countries, varying from border statistics (police and immigration) and supplemented by border surveys, to lodging establishments catered to tourism and standardized by \gls{unwto}. The dataset only considers overnight visitors, in that if a person from country $A$ is traveling to $B$ and takes a connecting flight in $C$ but does not stay there overnight, they will only count as a tourist in country $B$, and not in $C$. For ironing out statistical anomalies, we analyze an agglomerated set over all the years available from the data. This data also cover several years, allowing us to study the time-dependent properties throughout a number of years. The data for the \gls{wan} used in this analysis was provided by \textit{OpenFlights} \cite{openflights}. The geographical distance between countries' centroids (geographical center) is calculated by using the \textit{Haversine distance formula}, \cite{haversine}.

\begin{acknowledgments}
\textbf{Acknowledgments.} We acknowledge financial support from the Portuguese Foundation for Science and Technology (FCT) under Contract no. UID/FIS/00618/2019.
\end{acknowledgments}

\bibliographystyle{unsrt}
\bibliography{biblio}

\newpage

\section*{Supplementary Information} 
\label{supp}

\subsection*{Out-degree}
\begin{figure*}[htpb!]
\subfigure[Degree Distribution of the \gls{wtn}]{
\includegraphics[width=0.45\textwidth]{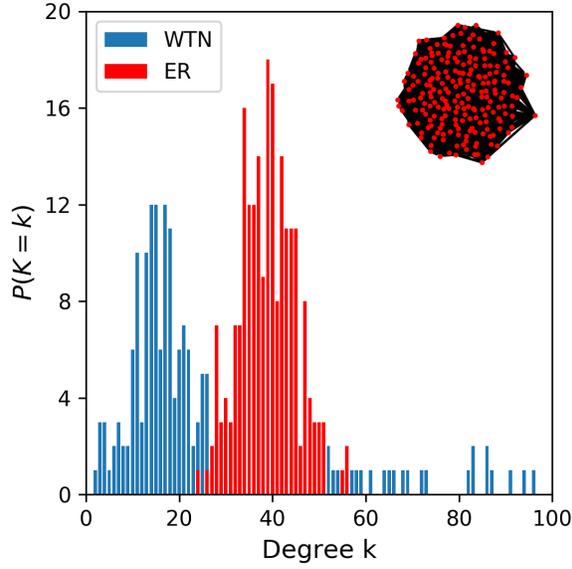}
\label{subfig:er}}
\subfigure[Cumulative Degree Distribution of the \gls{wan}]{
\includegraphics[width=0.45\textwidth]{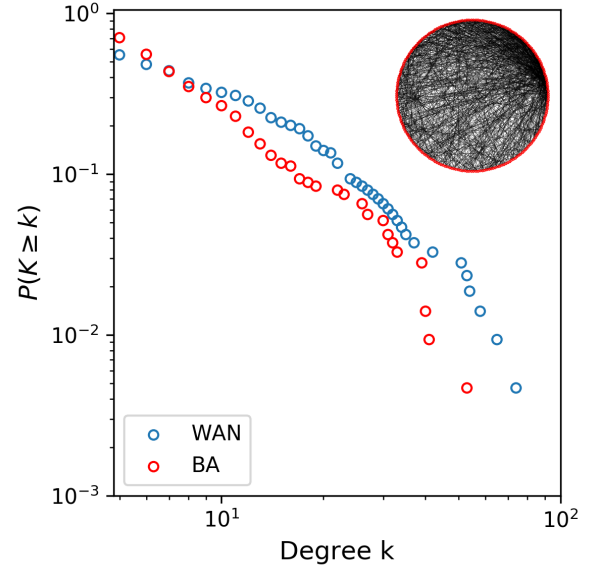}
\label{subfig:ba}}
\caption{\textbf{Comparison of the \gls{wtn} and the \gls{wan}}: (a) Degree distribution of the \gls{wtn} and an Erd\H{o}s\textendash R\'enyi model graph with the same number of nodes and with edge probability $p=0.91$, the same as for the \gls{wtn}. (b) Cumulative degree distribution of the \gls{wan} and a Barab\'asi\textendash Albert network with each node connected to $m=5$ other nodes.}
\label{fig:er_ba_image}
\end{figure*}

The average out-degree for the \gls{wtn} is $k^{out}_i = 19.38$ and its standard deviation has a value of  $\sigma_{<k^{out}>} = 11.67$. The distribution of the \textit{out-degree} parameter is show in Fig. \ref{fig:er_ba_image}, where we see a clear deviation from a power-law distribution, which would be expected for a real network like the \gls{wan} where the distribution is of power-law with an exponential cut-off \cite{trivikStructure}.

\subsection*{Triangle Decomposition}
The tables for the triangle distribution value and triangle participation coefficient for the \gls{wtn} and configuration model networks are shown in Table \ref{tbl:trianglesTable}. The triangle distribution indicates the frequency of each type of triangle in the network while the triangle participation tells us how many nodes belong in each type of triangle. The configuration model networks were built by randomizing the links present in the \gls{wtn} while keeping the degree distribution of its nodes intact. The results were obtained after simulating $100$ instances of these networks and taking the average of each parameter.

\begin{table}[htpb!]
\caption{Triangle distribution and participation values for the \gls{wtn} and configuration model networks. These networks were created by randomizing the links present in the \gls{wtn} while keeping the degree distribution of its nodes intact. The results present here are obtained after simulating $100$ instances of these networks and taking the average of each parameter.}
\label{tbl:trianglesTable}
\resizebox{0.475\textwidth}{!}{%
\begin{tabular}{cccccc}
                                                                                  &                                                       & Cycle         & In           & Out          & Bridge       \\ \cline{3-6} 
\multirow{2}{*}{\begin{tabular}[c]{@{}c@{}}Triangle\\ Distribution\end{tabular}}  & WTN*                                                  & 5.89\%        & 31.37\%      & 31.37\%      & 31.37\%      \\
                                                                                  & \begin{tabular}[c]{@{}c@{}}Conf.\\ Model\end{tabular} & 4.66(±0.29)\% & 31.78(0.1)\% & 31.78(0.1)\% & 31.78(0.1)\% \\ \hline
\multirow{2}{*}{\begin{tabular}[c]{@{}c@{}}Triangle\\ Participation\end{tabular}} & WTN*                                                  & 32.24\%       & 31.77\%      & 100\%        & 32.24\%      \\
                                                                                  & \begin{tabular}[c]{@{}c@{}}Conf.\\ Model\end{tabular} & 22.3(0.5)\%   & 22.3(0.5\%)  & 99.7(0.3)\%  & 22.9(0)\%   
\end{tabular}}
\end{table}

\subsection*{Alpha Parameter}
\begin{figure}[htpb!]
\includegraphics[width=0.5\textwidth]{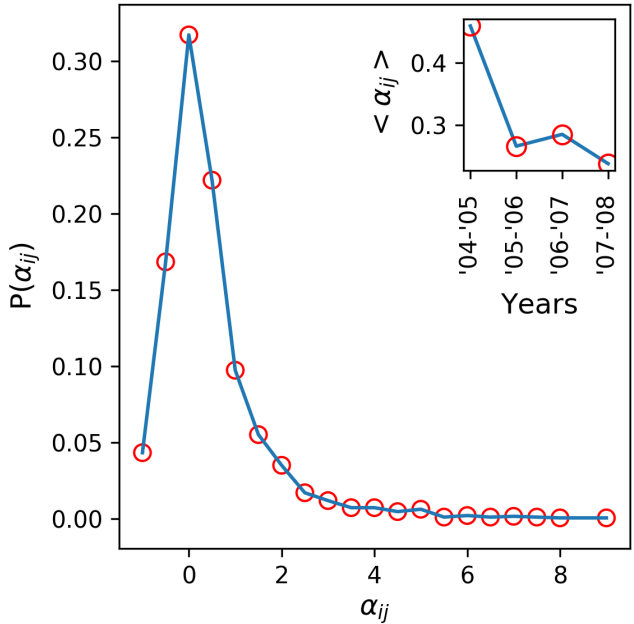}
\caption{\textbf{Change of Tourism}: Distribution of links in the \gls{wtn} in terms of their $\alpha$ parameter value with $t_0 = 2004$ and $t_n = 2008$. In the inset, we plot the $<\alpha_{ij}>$ over the years of the dataset.}
\label{pic:alphaParam}
\end{figure}

A simple measure of strength of evolution of tourism flow $\alpha_{ij}$ between a pair of countries $i$ and $j$ is defined as,
\begin{equation}
\alpha_{ij}(t_n) = \frac{F_{ij}^t(t_n) - F_{ij}^t(t_0)}{F_{ij}^t(t_0)},
\end{equation}
where $t_n$ and $t_0$ correspond to the end and start of discrete-events in the network dataset ($2008$ and $2004$, respectively), shows that big changes in flow have occurred for a very few connections over the years (trailing tail) and $\approx 30\%$ of the links have shown no change in traffic. Table \ref{tab:alphatopbot5} shows the difference in values for links that have evolved the most and least between $2004-08$ for which the dataset is available. Figure \ref{pic:alphaParam} shows the average alpha value spanning the entire dataset, $2004-2008$.

\begin{table}
\centering
\caption{Combination of countries representing representing change of tourism flow between them. The top half illustrates the links that have strengthened from $2004$ to $2008$. The bottom half are links that have weakened over time. The value measures is $\alpha$, a parameter to showcase the change in flow of tourism from $2004$ to $2008$ in the \gls{wtn}.}
  \begin{tabular}{ c | c | c }
    \multicolumn{1}{c}{Origin country} & \multicolumn{1}{c}{Destination country} & \multicolumn{1}{c}{$\alpha_{ij}$}\\
    \hline
    Macedonia & Israel & 881 \\
    Laos & Cambodia & 35.75\\ 
    Tajikistan & Kyrgyzstan & 35.23\\
    Rwanda & Ukraine & 26.11\\
    Uzbekistan & Kyrgyzstan & 17.66\\ 
    \hline \hline
	Kuwait & Israel & -0.93\\
	Georgia & Trinidad and Tobago & -0.96\\
	United Arab Emirates & Israel & -0.96\\
	Saudi Arabia & Israel & -0.99\\
	Macau & Malaysia & -0.99\\    	
  \end{tabular}
  \label{tab:alphatopbot5}
\end{table} 

\end{document}